\documentclass[%
 reprint,
superscriptaddress,
showpacs,preprintnumbers,
 amsmath,amssymb,
 aps,
]{revtex4-1}
\usepackage{soul}
\usepackage{ulem}
\usepackage{graphicx}
\usepackage{dcolumn}
\usepackage{bm}
\usepackage{color}
\newcommand\redout{\bgroup\markoverwith{\textcolor{red}{\rule[.5ex]{2pt}{0.4pt}}}\ULon}

\usepackage[colorlinks=true,citecolor=blue]{hyperref}
\hypersetup{colorlinks=true,citecolor=blue,linkcolor=red,urlcolor=blue}
\begin{document}
\date{\today}

\title{Exchange interaction of magnetic impurities in biased bilayer phosphorene nanoribbon}
\author{Moslem Zare }
\email{mzare@yu.ac.ir}
\author{Ebrahim Sadeghi}
\affiliation{Department of Physics, Yasouj University, Yasouj, Iran 75914-353, Iran}

\date{\today}

\begin{abstract}

 We study Ruderman-Kittel-Kasuya-Yosida (RKKY) interaction, in zigzag bilayer phosphorene nanoribbons (ZBLPNRs) under a perpendicular electric field.
 We evaluate the spatial and electric field dependency of static spin susceptibility in real space in various configurations of magnetic impurities at zero temperature.
 The electronic properties of ZBLPNR in the presence of a gate voltage is also obtained. In comparison to the other two-dimensional (2D) materials such as graphene, silicene and etc, ZBLPNR has two nearly degenerated quasiflat edge modes at the Fermi level, isolated from the bulk states. The band gap modulation of ZBLPNRs by the ribbon width and perpendicular electric field is investigated. Due to the existence of these quasiflat bands at the Fermi level, in the absence of the electric field, a sharp peak in the RKKY interaction is seen.
As shown, the signatures of these unique quasiflat edge modes in ZBLPNRs could be explored by using the RKKY interaction.
In the presence of a large bias potential a beating pattern of the RKKY oscillations occurs for when two magnetic impurities located inside the ZBLPNR. The electrically tunable RKKY interaction of ZBLPNRs is expected to have important consequences on the spintronic application of biased ZBLPNR.
\end{abstract}

\maketitle

\section{Introduction}

Black phosphorus (BP) monolayer, called phosphorene, has emerged as a promising material in the field of optoelectronics and magnetoelectronics of
2D systems ~\cite{Li14,Liu14,andre,Xia14,Moslem-bp,Carvalho16,Batmunkh}. Phosphorene has a puckered honeycomb structure due to the $sp^3$ hybridization~\cite{Liu15, Das14, Kamalakar15, Lu14, Fei14, Wang1, Moslem-bp}. Its conventional unit cell, consists of $4$ atoms with the lattice constants $a_x =3.3 $ \AA~ and $a_y =4.63 $ \AA~ in $x$ (zigzag) and $y$ (armchair) directions, respectively. Structural anisotropy of phosphorene makes its thermal, electrical and optical properties~\cite{zare-bp,Liu14,Fei14,Tran14}, a high degree of anisotropy. Charge carriers in phosphorene exhibit exceptionally high carrier mobilities at room temperature $\sim 1000$ cm$^2$ V$^ {-1} $ s$^{-1}$~\cite{Li14, Liu14} which exhibits a strongly anisotropic behavior in the phosphorene-based field effect transistor with a high on/off ratio, $\sim 10^4$~\cite{Liu14}. Moreover, by changing the number of layers and applying strain and external field, the direct bandgap of BP multilayer systems can be varied, spanning the losing gap between graphene and other 2D materials~\cite{Liu14, Du10, Tran14, cakir14, Liang14, Rodin14}. As the layer number strongly affects the physical properties of 2D black phosphorus multilayers, it is of both fundamental and practical interest to study the effect of the interlayer coupling on these physical properties. In this regard, bilayer phosphorene (BLP) is an appropriate candidate that can provide basic information on this coupling effect.

In the last decade, dilute semiconductors have emerged as a research hotspot due to their unique and new functionalities. In this regard, utilizing phosphorene may be lead to next generation of spintronic devices based on the spin degrees of freedom~\cite{fabian, Babar}.

Nonmagnetic nature of the pristine semiconductor phosphorene limits its applications in the field of magnetoelectronics and spintronics. Phosphorene can be magnetized in several ways such as doping with non-magnetic adatoms \cite{Zheng15,Yang16,khan15}, edge cutting~\cite{Zhu14,Du15,Farooq15,Farooq16,Xu16}, and inserting atomic defects and vacancies on phosphorene\cite{Srivastava15, Suvansinpan16, Babar}.
Phosphorene nanoribbons with bare zigzag edges are antiferromagnetic semiconductors \cite{Du15}. This magnetization also opens a significant direct band gap (about 0.7 eV), which transforms the metallic PNRs into semiconductors ones.

Most effective approach to induce magnetization with large Curie temperature is the adsorption or substitutional doping of $3d$ transition-metal atoms on phosphorene \cite{Zhai17, Khan16, Kulish, Seixas15, Sui15}. As the transition-metal atoms interact much stronger with phosphorene, compared to the other two dimensional materials such as graphene, it becomes more probable to turn the pristine phosphorene into a ferromagnetic or antiferromagnetic material \cite{Kulish}. Strength of these created magnetic moments depends on the metal species and the result can be tuned by the applied strain~\cite{Hu15, Sui15,Seixas15,Yu16,Cai17}.

Ruderman-Kittel-Kasuya-Yosida (RKKY) exchange interaction \cite{Ruderman, Kasuya, Yosida}, a fundamental interaction for spintronic applications, is mediated by a background of conduction electrons of the host material. It is the most important mechanism of the coupling between magnetic impurity dopants in metals and semiconductors. As an applied point of view, this interaction can lead a magnetically doped system to interesting phases such as ferromagnetic \cite{Vozmediano, Brey, Priour, Matsukura, Ko, Ohno-science}, antiferromagnetic \cite{Minamitani, Loss15}, spiral \cite{moslem-si, Mahroo-RKKY} and spin-glass \cite{pesin, Eggenkamp, Liu87}. Besides the practical importance of the RKKY interaction in the possible magnetic phases of semiconductor, it can provide information about the intrinsic properties of the material since this coupling is proportional to the spin susceptibility of the host system.
While this interaction falls off by $R^{-D}$, where $D$ is the dimension of the system \cite{fariborz-blg, fariborz-mos2}, it oscillates with the Fermi wavevector originates from the Friedel oscillations. In systems with multiband structure \cite{fariborz-mos2} or with spin polarization \cite{fariborz-sp}, these oscillations become more complicated than a monotonic oscillation with $\sin(2k_{\rm F}R)$ behavior, where $k_{\rm F}$ is the wave vector of the electrons (holes) at the Fermi level and $R$ is the distance of two magnetic impurities. Moreover, it has been shown that the magnitude of the RKKY interaction can be severely affected by the density of states (DOS) at the Fermi energy \cite{fariborz-blg, Mahroo-RKKY}. In addition, it can be sensitive to the direction of the distance vector between impurities in materials such as graphene \cite{sherafati-g, fariborz-blg} owing to the bipartite nature of the honeycomb sublattice. In materials with Rashba spin-orbit coupling, the spin response of the system to the magnetic impurity depends on the direction of the magnetic moment \cite{Mahroo-single} and as the result the RKKY interaction becomes anisotropic \cite{Mahroo-RKKY}.

RKKY interactions in nanoribbon of two dimensional lattices has attracted strong attention in condensed matter physics
\cite{moslem-si,Klinovaja13,Moslem18,Duan17}.
Recently, in a detailed study it has shown that the topological phase transition in the zigzag silicene nanoribbon can be probed by using the RKKY interaction \cite{moslem-si}. In another work, it has concluded that the RKKY interaction in the bulk phosphorene monolayer is highly anisotropic and the magnetic ground-state of two magnetic adatoms can be tuned by changing the spatial configuration of impurities as well as the chemical potential varying \cite{Moslem18}. Duan et.al also study the effect of strain on the magnetic impurity interactions in phosphorene \cite{Duan17}.
Very recently, the effect of tensile strain on the RKKY interaction in the biased monolayer phosphorene nanoribbon is studied, numerically \cite{Moldovan}.
They showed how one could isolate the edge state from that of the bulk contribution using the RKKY interaction, by tuning the external gate potential.
 In this paper, we present the Green's function technique for derivation of RKKY interaction in a biased bilayer phosphorene nanoribbon. Also, we show that by changing the perpendicular electric field due to the band structure changing one can drastically change the interaction between impurities, which can be a great way to control magnetic properties by electric field.

This paper is organized as follows. In Sec. \ref{sec:theory}, we introduce a tight-binding model Hamiltonian for biased bilayer phosphorene and then calculate the band spectrum of ZBLPNR under the vertical electric field. Here we introduce the theoretical framework which will be used in calculating the RKKY interaction from the real space Green’s function. In Sec. \ref{sec:results}, we discuss our numerical results for the proposed magnetic doped ZBLPNR in the presence of a perpendicular electric field. Finally, our conclusions are summarized in Sec. \ref{sec:summary}.

\begin{figure}[]
\begin{center}
\includegraphics[width=8.4cm]{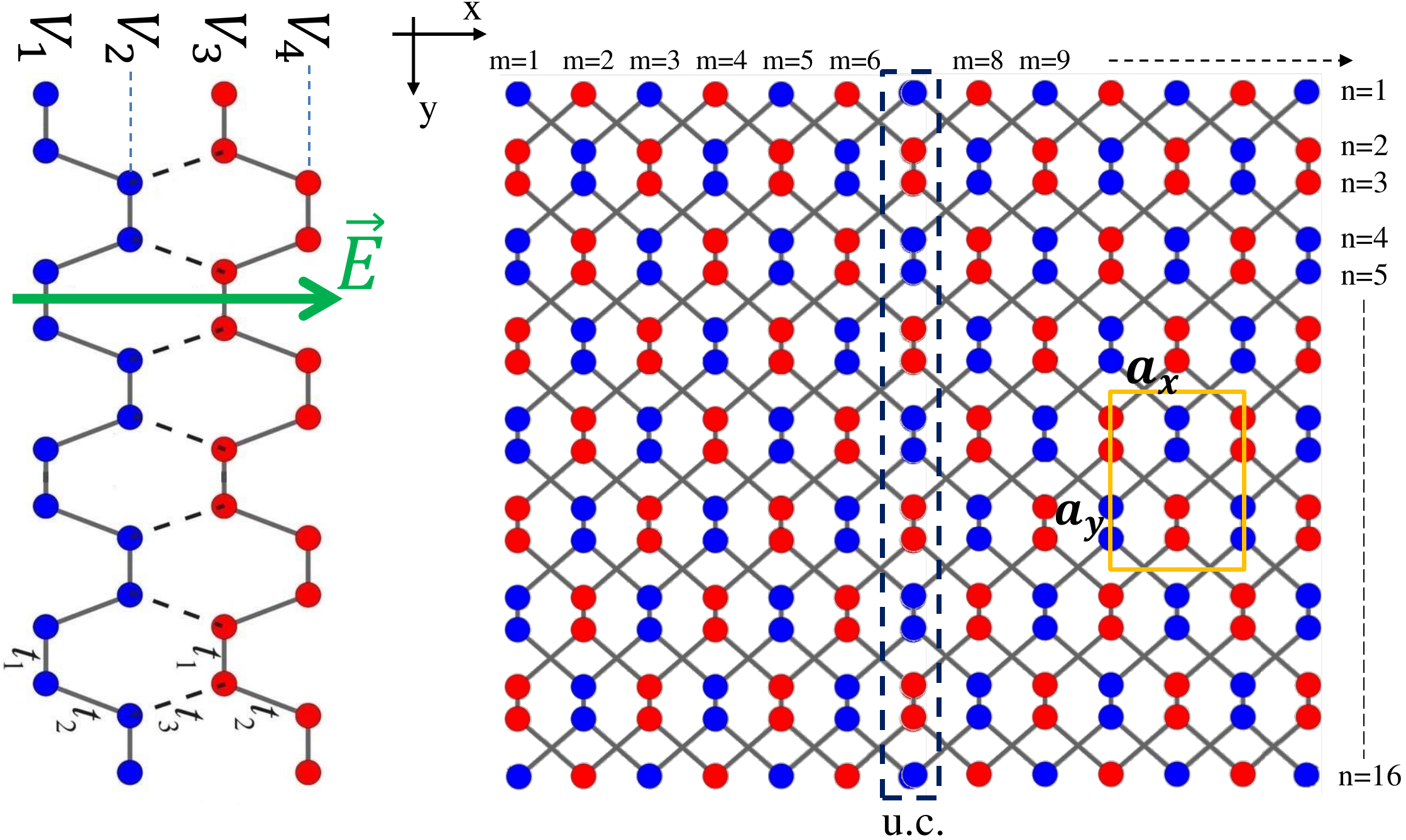}
\end{center}
\caption{\label{fig:1} (Color online)
 Sketch of the side view (left) and top view (right) of the two dimensional lattice structure of zigzag bilayer black phosphorus nanoribbon with $N=16$, in the presence of a perpendicular electric field $\vec{E}$. Blue and red circles correspond to atoms located in the bottom and top layers, respectively. The relevant hopping terms considered in Hamiltonian \ref{e1} are two in-plane hopping terms ($t_1=-1.21 ~eV,t_2=3.18~ eV$) and a $t_3=0.22~ eV$ interlayer term. The orange rectangle represents the unit cell of bilayer phosphorene with the lattice constants $a_x$ and $a_y$ and the dashed rectangle denotes the unit cell (u.c.) in the calculation of nanoribbon.}
\end{figure}

\begin{figure}[]
\begin{center}
\includegraphics[width=8.4cm]{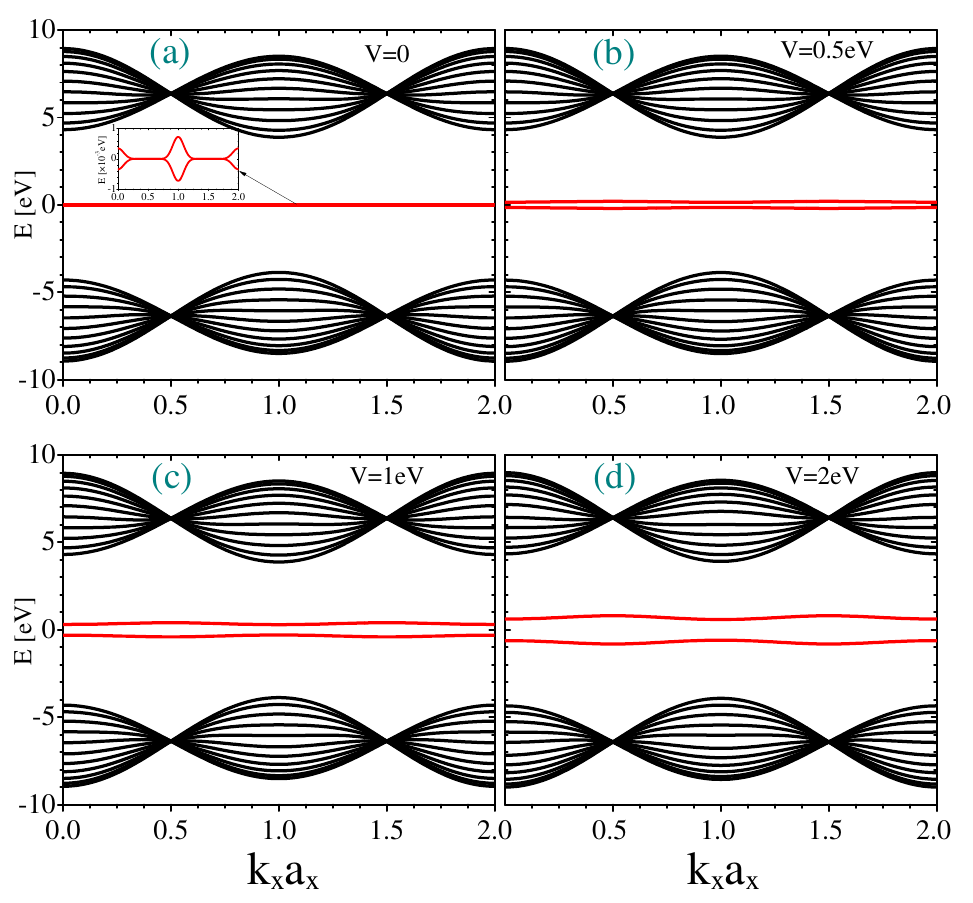}
\end{center}
\caption{\label{fig:2} (Color online) Energy bands for a ZBLPNR of $N=24$ with periodic boundary conditions in one direction ($x$) for several values of
external electric potential $V$, in which $k_x$ is
the wave vector parallel to the zigzag direction.}
\end{figure}

\begin{figure}[]
\begin{center}
\includegraphics[width=7.4cm]{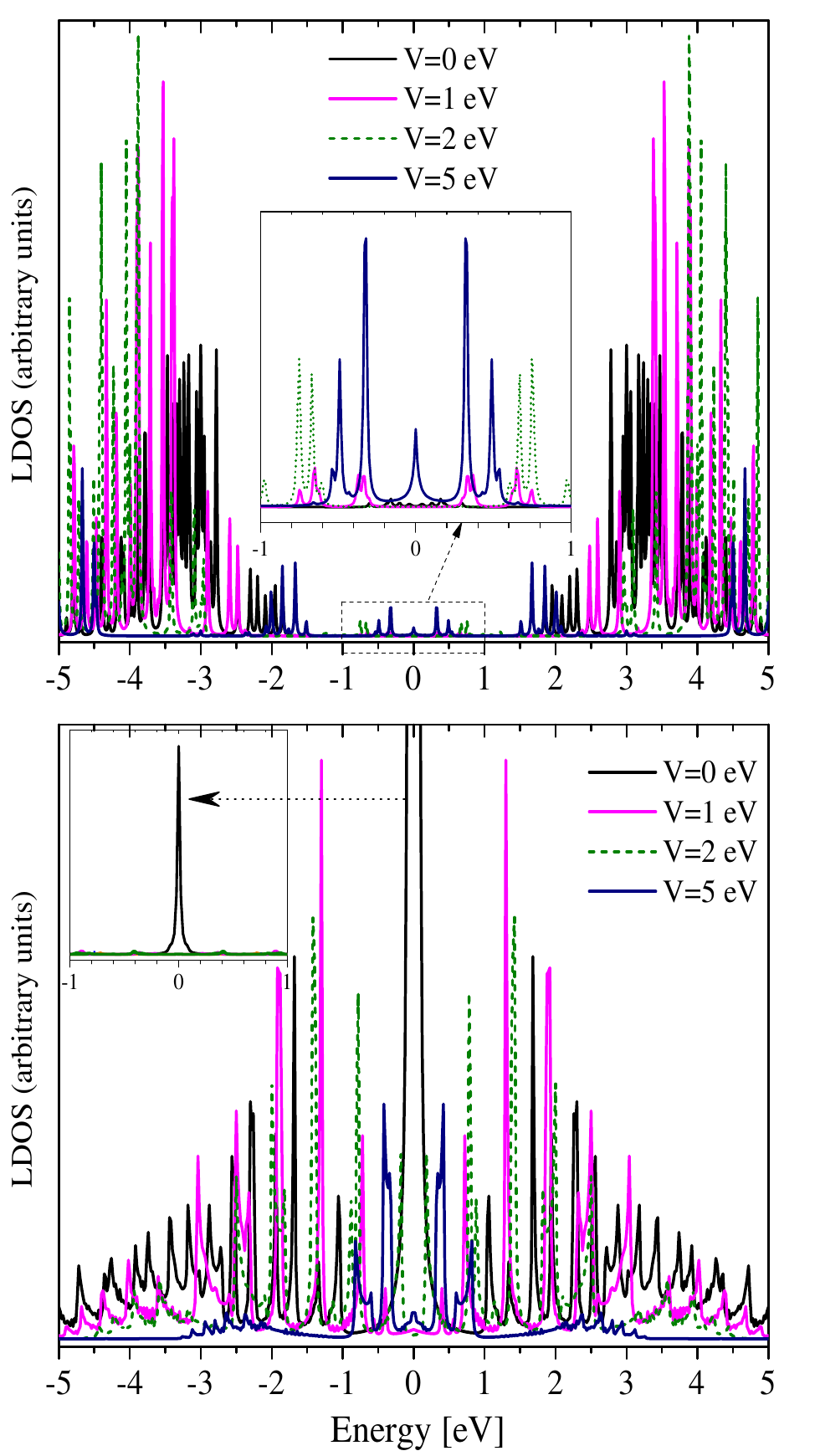}
\end{center}
\caption{\label{fig:dos} (Color online) Site-resolved local densities of states (LDOS) calculated for a ZBLPNR with $M=300,N=24$.
Top: for a bulk site with coordinate $(150,12)$ and bottom: for an edge site with $(150,1)$.
For the better clarity, the LDOS within the energy range of $-1$ to $1~eV$ is shown in the insets. When the applied gate voltage is zero ($V=0$) the LDOS shows a sharp peak around the zero energy corresponding to the edge states in comparison to the bulk (see insets).}
\end{figure}

\begin{figure}[]
\begin{center}
\includegraphics[width=9.8cm]{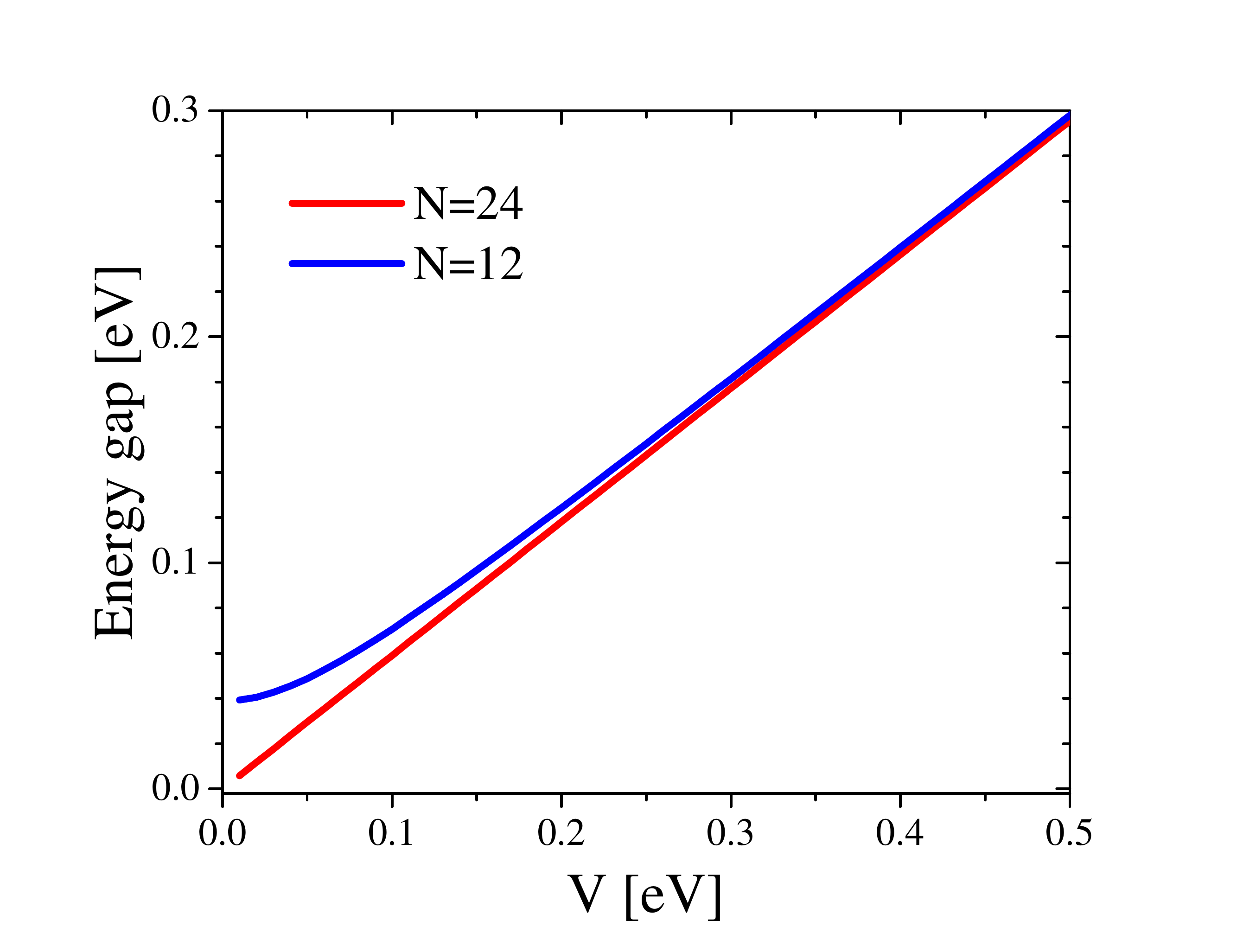}
\end{center}
\caption{\label{fig:3} (Color online) Energy gap of bilayer phosphorene nanoribbons with widths $N=12, 24$ as a
function of the perpendicular electric potential.}
\end{figure}

\section{Theory and model}\label{sec:theory}

 As the Bernal stacking configuration of two phosphorene layers, coupled via the van der Waals interaction, is energetically most stable form of the bilayer phosphorene \cite{Peeters15,Jhun17}, we consider an AB stacked ZBLPNR as shown in Fig. \ref{fig:1}. In the presence of an uniform perpendicular electric field, low-energy carriers in BLP are described by the following tight-binding(TB) Hamiltonian~\cite{Moldovan,Rudenko2016,Dolui16}:
\begin{equation}\label{e1}
H=\sum_{i}V_ic_i^{\dag}c_i+\sum_{i\neq j}t_{ij}^{\|}c_i^{\dag}c_j+\sum_{i\neq j}t_{ij}^{\perp}c_i^{\dag}c_j,
\end{equation}
where the summation runs over all lattice sites of the system. $c_i^{\dag}$ ($c_j$) is the creation (annihilation) operator of an electron at site $i$ ($j$),
$V_i$ is the on-site energy at site $i$ and $t_{ij}^{\|}$ ($t_{ij}^{\perp}$) is the intralayer (interlayer) hopping energy between sites $i$ and $j$. The relevant hopping parameters considered by Li et.al \cite{Moldovan}, are two in-plane hopping terms ($t_1=-1.21~ eV,t_2=3.18~ eV$) and a $t_3=0.22~ eV$ interlayer term. In the presence of a perpendicular electric field, the four atomic sublayers in BLP will earn different on-site electrostatic potentials in the form of $V_1=(1/2+\epsilon)V$, $V_2=(1/2-\epsilon)V$, $V_3=(-1/2+\epsilon)V$, and $V_4=(-1/2-\epsilon)V$, where $V=eEd$ is the electrostatic potential energy difference between the top and bottom phosphorene layers, with $e$ the elementary charge, $E$ the electric field strength, $d$ the interlayer separation, and $\epsilon=0.202$ is a linear scaling factor that accounts for the sublayer dependence of the on-site electrostatic potential \cite{Yuan16}.

The geometry of a ZBLPNR with zigzag edges is illustrated in Fig.\ref{fig:1}. Its conventional unit cell, consists of $8$ atoms with the lattice constants $a_x$ and $a_y$ in $x$ (zigzag) and $y$ (armchair) directions, respectively exactly equal to the monolayer phosphorene lattice constants. Here, the unit cell used in the tight-binding calculations of the ZBLPNR (dashed rectangle), containing $N$ atoms, is also indicated. The respective unit cell width is $a_x$. For simplicity, as shown in this figure each atom labeled with a set $(m,n)$, which $m,n$ represent the $x$ and $y$ coordinates of the lattice points. In our analysis, we consider the two magnetic impurities located at $(m_1,n_1)$ and $(m_2,n_2)$ sites of the nanoribbon (following the notations of Fig.\ref{fig:1}). In this ribbon geometry, it is easy to find the energy dispersion with the periodic boundary condition along the ribbon's zigzag edge in the $x-$direction. Owing to the translational invariant along the ribbon edges $(x)$, the momentum in the $x-$direction is a good quantum number. To study the band structure properties provided by our tight-binding model, we find its ${\bf k}-$space forms as
$\sum_{{\bf k}}\psi_{{\bf k}}^{\dag} H_k \psi_{{\bf k}}$. Applying Bloch’s theorem, performing the Fourier transformation along the $x-$direction, the
Hamiltonian in ${\bf k}-$space can be written as

\begin{equation}\label{eq:H-k1}
 H_k=H_{AA}+H_{BB}+H_{AB}e^{-i k_x
a_x}+H_{AB}^\dagger e^{i k_x a_x}
\end{equation}
in which $a_x$ is the lattice constant along the zigzag direction. Moreover, $H_{AA}$ and $H_{BB}$ describe coupling matrix within the principal unit cells (intralayer), with odd and even indices $n$, respectively and $H_{AB}$ denotes the effective coupling between two adjacent unit cells (interlayer), based on the tight-binding model given by Eq.~(\ref{e1}).

The energy dispersion and the wavefunctions can also be obtained by diagonalizing the Hamiltonian for the nearest-neighbor tight-binding model.
In order to calculate the site-resolved local density of states (LDOS) for the ribbon as $ \rho_i({\bf r},E)=-\frac{1}{\pi}{\rm Im}[G_{ii}({\bf r,r,}E)]$, for $i$-th site. In order to do so, corresponding wave function for a given energy and wave vector might be first obtained.

The calculated band structures of ZPNRs are shown in Figs.\ref{fig:2}. In similarity with ZPNR, ZBLPNRs have two nearly degenerated quasiflat edge modes at the Fermi level \cite{Ajanta16,Ezawa14,Taghizadeh15,Ma16,Ostahie16} that are entirely detached from the bulk band. By applying the electric field the band degeneracy breaks.
The properties of edge states in ZBLPNRs are essentially different from the other 2D zigzag nanoribbons.
In comparison with ZBLPNR and ZPNR, in 2D Dirac materials such as graphene and silicene, the edge modes merge into the bulk band at the two Dirac points.
Degeneracy of these two quasi flat bands is broken by applying a perpendicular electric field. As recently addressed by Ezawa \cite{Ezawa14}, the origin of this decoupled matter of the flat edge modes is the presence of two out-of plane zigzag chains, coupled by a relatively strong hopping parameter.

Figure \ref{fig:dos} displays the LDOS of phosphorus atoms in either sublattice at the edge or in the bulk. The top panel presents the LDOS of the bulk configuration at the site $(150,12)$, and the bottom one presents the edge configuration at the site $(150,1)$. Similar to the ZPNR \cite{Firoz18}, when the applied gate voltage is zero the LDOS shows a sharp peak around the zero energy corresponding to edge states, in comparison to the bulk.

Moreover, figure \ref{fig:3} shows the energy gap $E_g$ of the ZBLPNR as a function of the applied electric field. As previously obtained, the energy gap increases linearly as the applied electric field is increased \cite{Yang16}.
The zero biased gap decreases with increasing ribbon width and tends to zero in the limit of very large $N$.
The carrier-mediated exchange coupling between the spin of itinerant electrons and two magnetic impurities with magnetic moments ${\bf S}_1$ and ${\bf S}_2$, located respectively at ${\bf r}$ and ${\bf r'}$, is given by

\begin{equation}
V = - \lambda \ ( {\bf S}_1 \cdot {\bf s(r)} + {\bf S}_2 \cdot {\bf s(r')} ),
\label{hamilint}
\end{equation}
where ${\bf s(r)},{\bf s(r')} $ are the conduction electron spin densities at positions ${\bf r}$ and ${\bf r'}$ and $\lambda$ is the coupling between the impurity spins and the itinerant carriers.

In the linear response regime, the interaction energy between the two localized magnetic moments may be written as a Heisenberg form~\cite{Ruderman,Kasuya,Yosida,Imamura}
\begin{equation}
E({\bf r},{\bf r'}) = J ({\bf r},{\bf r'}) {\bf S}_1 \cdot {\bf S}_2,
\label{RKKYE}
\end{equation}

The RKKY interaction $J ({\bf r},{\bf r'}) $ is explained using the susceptibility, the response of the charge density $n$ to a perturbing potential $V$,
\begin{equation}
J ({\bf r},{\bf r'}) = \frac{\lambda^2 \hbar^2 }{4} \chi ({\bf r},{\bf r'}).
\label{RKKYJ}
\end{equation}

where $ \chi({\bf r},{\bf r'}) \equiv  \delta n({\bf r}) / \delta V({\bf r'})$ is the charge susceptibility for a crystal, $\delta V(r')$ is a spin-independent perturbing potential and $\delta n({\bf r})$ is the induced charge density.

The static spin susceptibility can be written in terms of the integral over the unperturbed Green's function
\begin{equation}
\chi ({\bf r},{\bf r'}) =
- \frac{2}{\pi} \int^{\varepsilon_F}_{-\infty} d\varepsilon \
{\rm Im} [G^0 ({\bf r}, {\bf r'}, \varepsilon) G^0 ({\bf r'},{\bf r}, \varepsilon)],
\label{chiGG}
\end{equation}
where $\varepsilon_F$ is the Fermi energy. The expression for the susceptibility may be obtained by using the spectral representation of the Green's function
\begin{equation}
G^0 ({\bf r},{\bf r'},\varepsilon)= \sum_{n,s} \frac{\psi_{n,s}({\bf r})\psi^{*}_{n,s}({\bf r'})}{\varepsilon+i\eta - \varepsilon_{n,s}},
\label{GFspct}
\end{equation}
where $\psi_{n,s}$ is the sublattice component of the unperturbed eigenfunction with the corresponding energy $\varepsilon_{n,s}$. For a crystalline structure, ${n,s}$ denotes the band index and spin. In other words, it just denotes a complete set of quantum states. Substituting Eq. (\ref{GFspct}) into Eq. (\ref{chiGG}), we get the result
\begin{widetext}
\begin{eqnarray}
\chi ({\bf r},{\bf r'}) =- \frac{2}{\pi} \int^{\varepsilon_F}_{-\infty} d\varepsilon &\times& \nonumber \\
 \sum_{\substack{{n,s} \\{ n',s'}}}  \{  {\rm Re} \ [\psi_{n,s}({\bf r})\psi^{* }_{n,s}({\bf r'})   \psi_{n',s'}({\bf r'})\psi^{* }_{n',s'}({\bf r})] \ &{\rm Im}& \ [(\varepsilon+i\eta -\varepsilon_{n,s})(\varepsilon+i\eta -\varepsilon_{n',s'}) ]^{-1}
 \nonumber \\
+ {\rm Im} \ [\psi_{n,s}({\bf r})\psi^{* }_{n,s}({\bf r'})\psi_{n',s'}({\bf r'})\psi^{*}_{n',s'}({\bf r})] \  &{\rm Re}& \ [(\varepsilon+i\eta -\varepsilon_{n,s})(\varepsilon+i\eta -\varepsilon_{n',s'}) ]^{-1} \}.
\end{eqnarray}
\end{widetext}

After exchanging the dummy variables $n,s$ and $ n',s'$, the imaginary part of the product of the four wave functions appearing in the equation above is odd, while its real part is even, and at the same time, both the imaginary and the real parts of the product of the energy-space Green's function are even. By applying this property the second expression becomes zero. Finally, the $\chi({\bf r},{\bf r'})$ reads as
\begin{equation}
\chi({\bf r},{\bf r'}) =\sum_{\substack{{n,s} \\{n',s'}}}
\psi_{n,s}({\bf r})\psi^{*}_{n,s}({\bf r'})\psi_{n',s'}({\bf r'})\psi^{*}_{n',s'}({\bf r})  \
\mathcal{E}(n,s,n',s'),
\label{chiE1}
\end{equation}
in which
\begin{eqnarray}
\mathcal{E}= 2 \int_{-\infty}^{\varepsilon_F} d\varepsilon \ \big[ \frac{\delta(\varepsilon -\varepsilon_{n',s'})}  {\varepsilon -\varepsilon_{n, s}} +  \frac{\delta(\varepsilon-\varepsilon_{n,s})}  {\varepsilon-\varepsilon_{n',s'}}  \big ] .
\label{chikk'}
\end{eqnarray}

To prove the above equality we use the relationship $\lim_{\eta \rightarrow 0^+} (x\pm i \eta)^{-1}=\mathcal{P}(1/x) \mp i \pi \delta(x)$.
The integration over energy can be carried out next, leading to our desired result
\begin{eqnarray}
\chi({\bf r},{\bf r'}) &&=2 \sum_{\substack{n,,s \\ n',s'}}[ \frac{f(\varepsilon_{n,s})-f(\varepsilon_{n',s'})}{\varepsilon_{n,s}-\varepsilon_{n',s'}}\nonumber\\
&&\times \psi_{n,s}({\bf r})\psi^{* }_{n,s}({\bf r'})\psi_{n',s'}({\bf r'})\psi^{*}_{{ n'}s'}({\bf r})].
\label{chiE}
\end{eqnarray}
where, $f(\varepsilon)$, is the Fermi function.
This is a well-known formula in the linear response theory that is the main equation in this work. It is worth mentioning that under the interchange of $n,s$, and $ n',s'$, the summand in Eq. (\ref{chiE}) converts to its complex conjugate, so that only the real part survives.

\section{numerical results}\label{sec:results}
In this section, we present our main results for the RKKY exchange coupling in the zigzag bilayer phosphorene nanoribbons presence of perpendicular electric field. In order to do so, we evaluate static spin susceptibility Eq. (\ref{chiE}), in real space in various configurations of magnetic impurities at various bias voltages.

\begin{figure}[]
\begin{center}
\includegraphics[width=9.4cm]{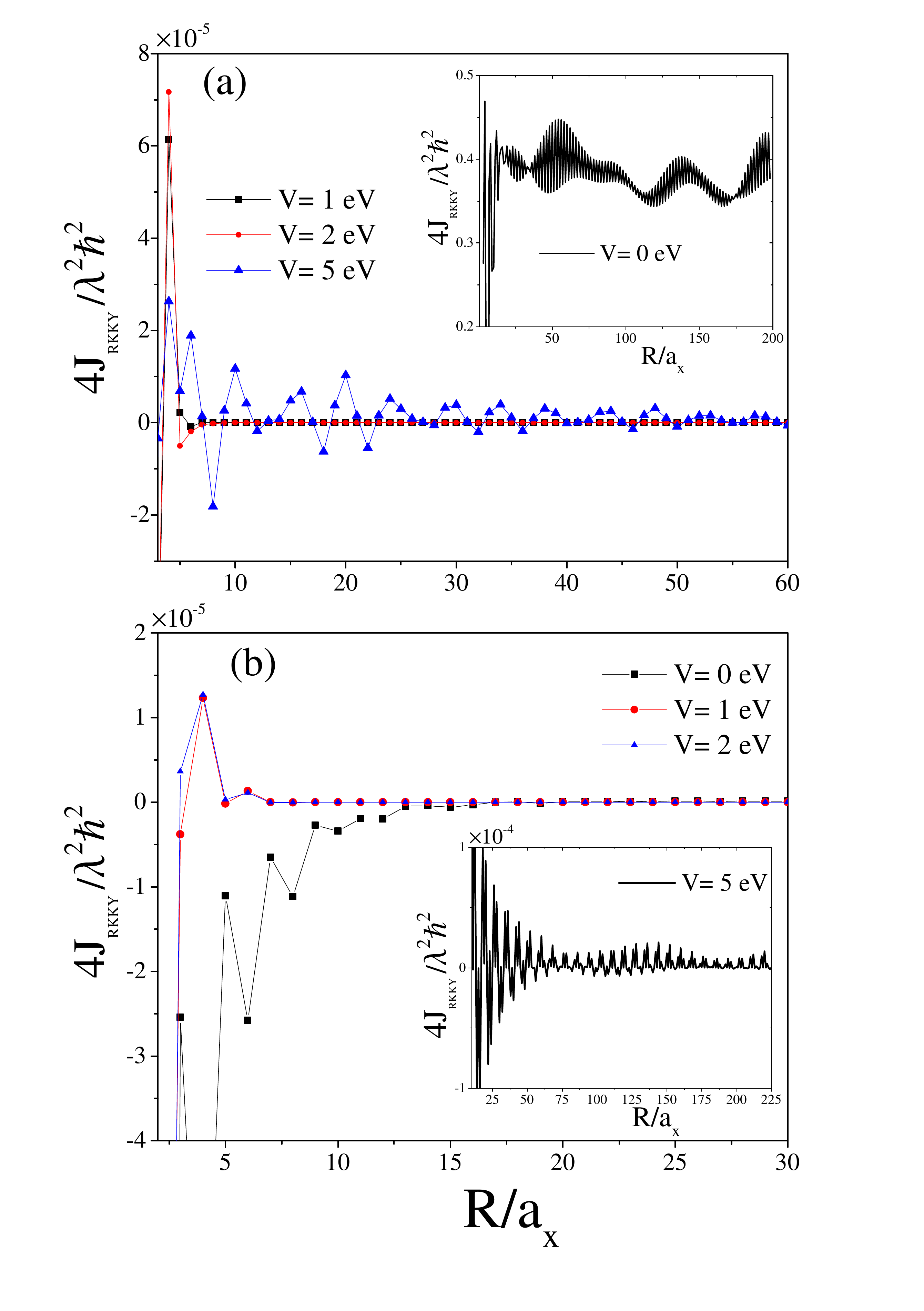}
\end{center}
\caption{\label{fig:4} (Color online) Scaled RKKY interaction as a function of the impurities distance for the ZBLPNR with $M=300,N=24$. (a) impurities located on the same edge such that the first impurity fixed at the sublattice $(10,1)$ and the second one is located at (m,1) where $m=11,12, 13,...$. As in the absence of the electric potential, the strength of the RKKY interaction is approximately five orders of magnitude larger than the biased ZBLPNR, we show the result for $V=0$ in the inset. (b) impurities located inside the ZBLPNR such that the first impurity located on the sublattice $(10,12)$ and the second one is placed at $(m,12)$ with $m=11,12, 13,...$ for various electric potentials. For the better clarity, we show the result for $V=5~eV$ in the inset.  }
\end{figure}

\begin{figure}[]
\begin{center}
\includegraphics[width=8.6cm]{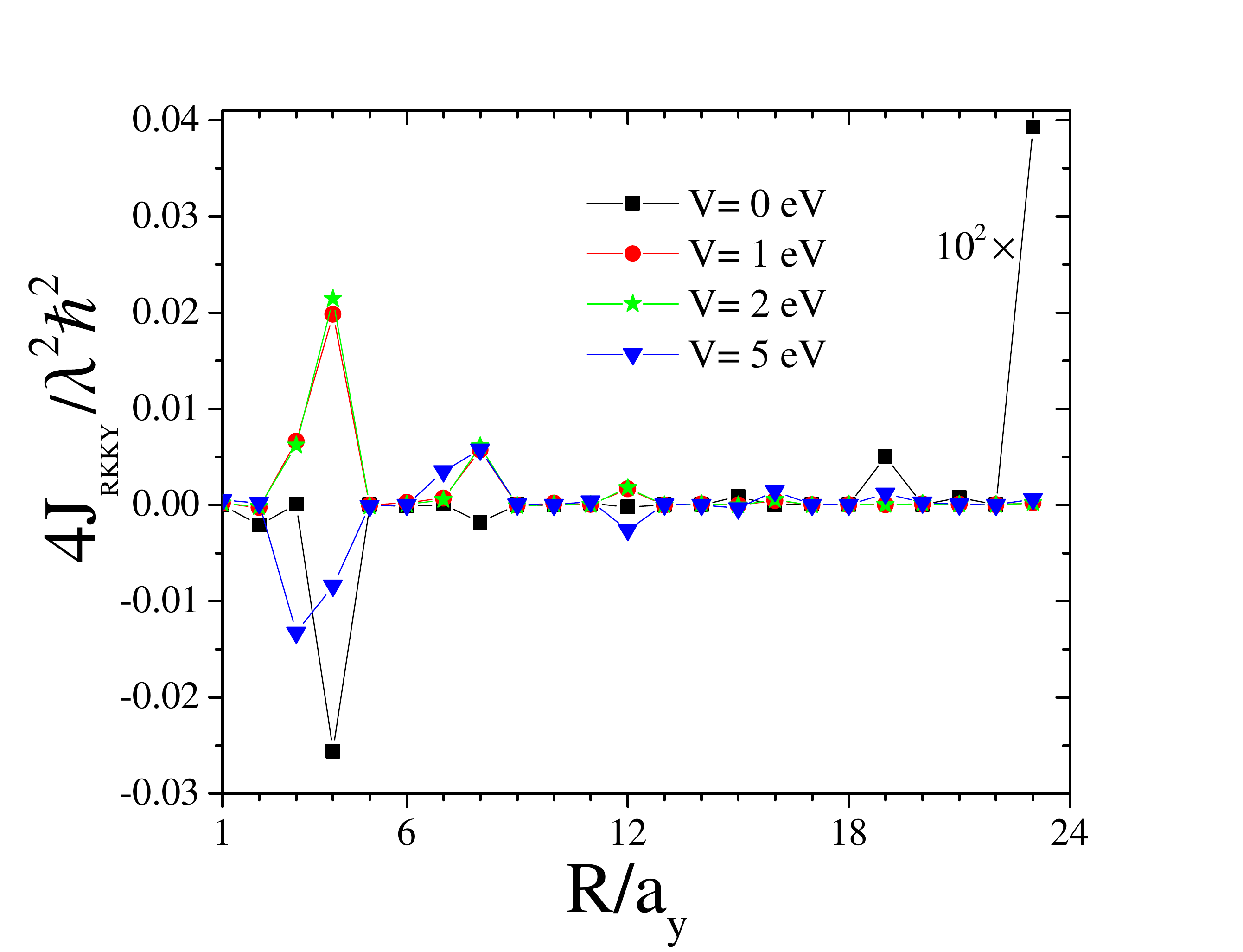}
\end{center}
\caption{\label{fig:5} (Color online) Scaled RKKY interaction as a function of the impurities distance for the ZBLPNR with $M=300,N=24$, when the first impurity is fixed at the edge at sublattice $(150,1)$ and the second one is moved along the armchair line $m=150$ with coordinates $(150,n)$, with $n=2,3,...24$.}
\end{figure}

Figure \ref{fig:4} shows the spatial behavior of the RKKY interaction for two impurities, (a) both sitting on the same edge along the line $n=1$ (b) both located inside the zigzag nanoribbon (away from the edges) along the line $n=12$ for different bias voltages for a ZBLPNR with $M=300, N=24$. As indicated, in the case of both impurities located at an edge, in the moderate fields, the RKKY interaction displays an oscillatory behavior in $R$ and decays fast with a short-ranged behavior. More interestingly, by applying the electric field the RKKY interaction strength dramatically quenched (in the zero field regime ($V=0$), the RKKY interaction is about five orders of magnitude greater). The reason for that is related to the existence of nearly-zero-energy states at the edge of the ZBLPNR. In the high field regime ($V=5 ~eV$), the RKKY interaction shows a long-range oscillatory behavior in $R$, because the Fermi energy crosses the bulk bands.
Similar to the ZPNR~\cite{Firoz18}, the RKKY coupling attains its maximum strength when the applied gate voltage is zero.
The contribution of the bulk states are nearly zero as the impurities are located at the edge of the ZBLPNR.

Otherwise, when both the impurities are sited in the interior region of the ZBLPNR, no difference in the order of magnitude appears for different bias voltages (see figure \ref{fig:4}(b)). In similarly with the edge configuration of the impurities (both located on the same edge), RKKY shows a few oscillations in $R$, then it decays fast with a short-ranged behavior.
More interestingly, in the presence of a large bias potential ($V=5 ~eV$), a beating pattern of oscillations of the RKKY interaction occurs for when two magnetic impurities located inside the ZBLPNR. The beating feature originated from the crossing the Fermi energy and the bulk bands for the bulk impurities (both moments located inside the ZBLPNR).

Figure \ref{fig:5} shows the spatial behavior of the RKKY interaction for when the first impurity is fixed at the edge at sublattice $(150,1)$ and the second one is moved along the line $m=150$ with coordinates $(150,n)$, where $n=2,3,...24$. As in the absence of the applied bias, ($V=0$) the exchange coupling is approximately two orders of magnitude greater than the case with potential ($V\neq0$), for the better clarity we multiply the values related to the zero bias by $10^{-2}$.
In the case of zero bias, for both impurities located at an armchair chain (both located in the same ZBLPNR unit cell), the largest coupling appear when both moments is located at two counterpart edge sublattices.

\begin{figure}[]
\begin{center}
\includegraphics[width=8.4cm]{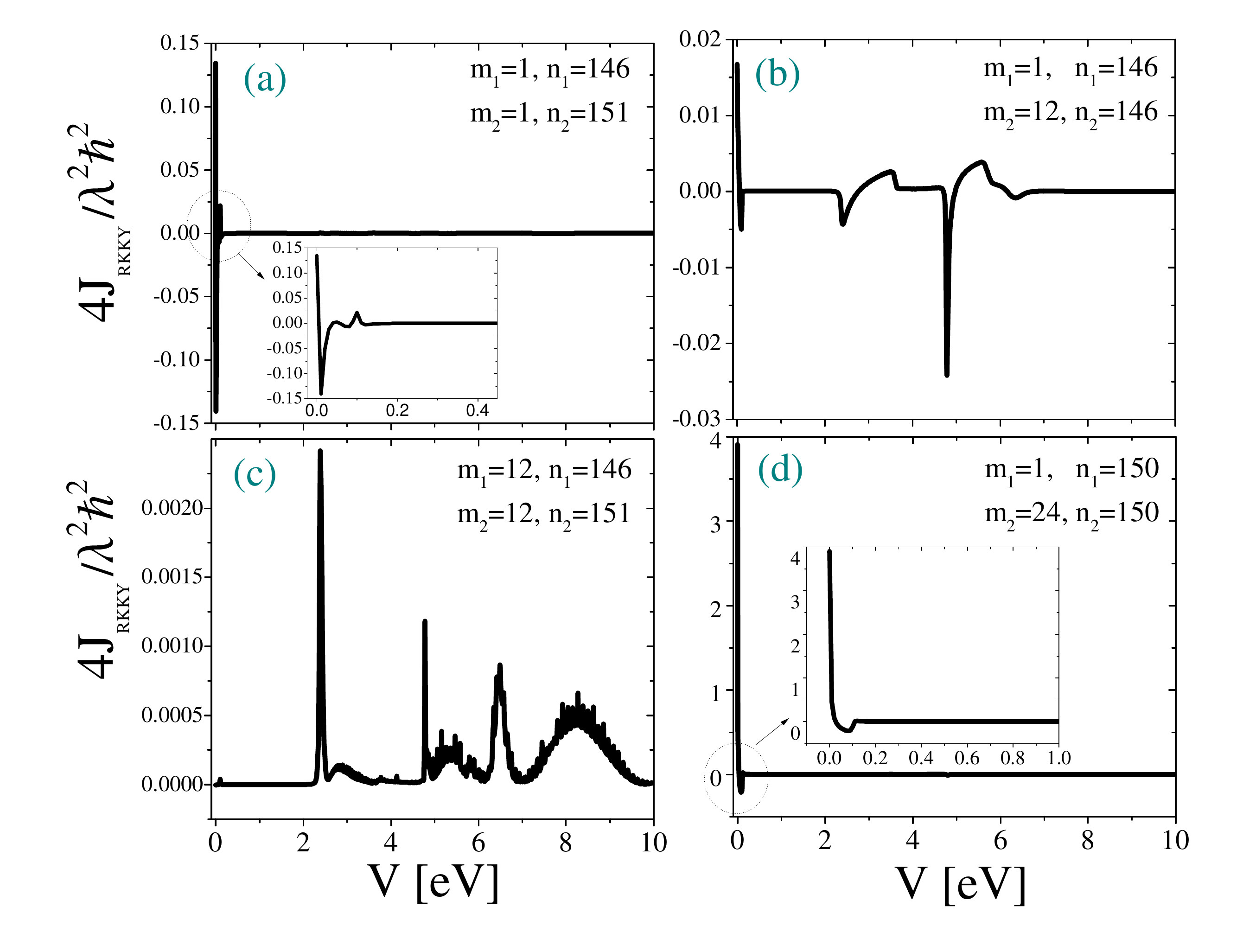}
\end{center}
\caption{\label{fig:6} (Color online) Scaled RKKY interaction as a function of the perpendicular electric potential for $M=300,N=24$, when (a) both impurities are fixed at the zigzag edge at $(146,1)$ and $(151,1)$ lattice points (b) the first impurity is fixed at the zigzag edge at $(150,1)$ and the second one located inside the ZBLPNR at $(150,12)$ lattice point (c) both impurities located inside the ZBLPNR at $(146,12)$ and $(151,12)$ lattice points (d)impurities are fixed at the counterpart zigzag edges at $(146,1)$ and $(146,24)$ lattice points.}
\end{figure}

Figures \ref{fig:6}(a-d) show the numerical results for the RKKY coupling as a function of bias voltage for different configurations of magnetic impurities. (a) when two impurities are fixed at the same zigzag edge at $(148,1)$ and $(153,1)$ lattice points. Turning on the electric field, a sudden fall off in the RKKY interaction strength occurs, because the zero-energy state at the edge of the ZBLPNR is suddenly dropped. This behavior is also seen for when both moments are located at two counterpart edge sublattices (figure \ref{fig:6}(d)) because by turning on the electric field, the edge modes in ZBLPNR become fully separated from the bulk. This unique nature of the edge states in ZBLPNR allows us to probe them separately from the bulk.
Numerical results for when the first impurity is fixed at the zigzag edge at $(150,1)$ and the second one located inside the ZBLPNR at $(150,12)$ lattice point, is shown in figure \ref{fig:6}(b). Figure \ref{fig:6}(c) shows the RKKY coupling for two impurities fixed inside the ZBLPNR at $(146,12)$ and $(151,12)$  lattice points.

\section{summary}\label{sec:summary}.

 We study Ruderman-Kittel-Kasuya-Yosida interaction, in zigzag bilayer phosphorene nanoribbons (ZBLPNRs) under a perpendicular electric field.
 For this, we evaluate the spatial and electric field dependency of static spin susceptibility in real space in various configurations of magnetic impurities at zero temperature. We evaluate this interaction from the Green's function approach based on the tight-binding model Hamiltonian.
 The electrically tunable RKKY interaction of ZBPNRs is expected to have important consequences on the spintronic application of biased ZBLPNRs.
 The electronic property of ZBLPNR in the presence of gate voltage is also obtained. In comparison to the other 2D materials such as graphene, silicene and etc, ZBLPNRs have two nearly degenerated quasiflat edge modes at the Fermi level in a ZBLPNR, isolated from the bulk states. A gap occurs by applying a perpendicular electric field. As shown, the signatures of these unique quasiflat edge modes in ZBLPNRs could be explored by using the RKKY interaction.
 Due to the existence of the quasiflat bands at the Fermi level, in the absence of the electric field, a sharp peak in the RKKY interaction is seen. By varying the width of the phosphorene ribbons, we find that the size effect is crucial for determining the relative importance of the edge state.Moreover, figure \ref{fig:3} shows the energy gap $E_g$ of the ZBLPNR as a function of the applied electric field. As previously obtained, the energy gap increases linearly as the applied electric field is increased \cite{Yang16}.
The zero bias gap decreases with increasing ribbon width and tends to zero in the limit of very large $N$.
In the presence of a large bias potential a beating pattern of oscillations of the RKKY interaction occurs for when two magnetic impurities located inside the ZBLPNR.
\section*{ACKNOWLEDGMENT}

This work is partially supported by Iran Science Elites Federation.

\end{document}